\documentclass{article}
\usepackage{graphicx} 
\usepackage{amsmath}
\usepackage{float} 
\usepackage[indent=12pt]{parskip} 
\usepackage{fewerfloatpages} 
\usepackage{authblk} 
\usepackage{cleveref}
\usepackage{array} 
\usepackage{booktabs} 
\usepackage{placeins} 

\usepackage[
backend=biber,
style=apa,
sorting=nyt
]{biblatex}
\title{Nutrient Competition as a General Mechanism of Regulation in Photosymbiosis}
\author[1,2,3,${\ast}$,${\P}$]{Jordan A. Gault}
\author[1,3,${\dagger}$]{Nils Rädecker}
\author[1,2,3,${\ddagger}$]{Iliana B. Baums}
\author[1,2,3,${\S}$]{Thilo Gross}
\affil[1]{HIFMB, Helmholtz Institute for Functional Marine Biodiversity, Oldenburg, Germany}
\affil[2]{University of Oldenburg, Institute for Chemistry and Biology of the Marine Environment, Oldenburg, Germany}
\affil[3]{Alfred-Wegener Institute, Helmholtz Center for Polar and Marine Research, Bremerhaven, Germany.}
\affil[${\ast}$]{Corresponding Author: jordan.gault@uni-oldenburg.de}
\affil[${\P}$]{Im Technologiepark 5, Oldenburg 26129}
\affil[${\dagger}$]{nils.raedecker@hifmb.de}
\affil[${\ddagger}$]{iliana.baums@hifmb.de}
\affil[${\S}$]{thilo2gross@gmail.com}
\date{}

\begin{document}

\maketitle

\begin{abstract}
    Nitrogen limitation is a key mechanism regulating the mutualism between photosymbiotic cnidarians and their algal symbionts. Algal-derived photosynthate facilitates host assimilation of inorganic nitrogen, specifically ammonium. The resulting competition between symbiotic partners for inorganic nitrogen leads to a negative feedback loop wherein symbiont density is effectively controlled through facilitation of their competitor. However, it is unclear whether this feedback loop is sufficient to stabilize the interaction absent other forms of regulation. Using a generalized modeling approach, we identify the conditions under which cnidarian-algal competition for nitrogen yields dynamically stable symbioses. We show that nitrogen limitation is sufficient to stabilize the interaction through the interplay of intraspecific (symbiont-symbiont) and interspecific (host-symbiont) competition. As host assimilation of inorganic nitrogen increases, the system shifts from being stabilized by intraspecific competition between symbionts to interspecific competition between host and symbionts, necessitating a positive coupling of symbiont density to host nitrogen assimilation via photosynthate translocation for dynamic stability. 
\end{abstract}

\clearpage

\section{Introduction}

Mutualisms initially posed an ecological puzzle: mutually reinforcing positive feedbacks were thought to make the interaction ecologically unstable \autocite{hale2021ecological, holland2015population}, leading to an ``orgy of mutual benefication" \autocite{may1976models}. Progress has since been made in understanding how various forms of negative density dependence can stabilize mutualistic interactions over ecological timescales \autocite{hale2021ecological, holland2015population, massing2025generalized}. Such negative density dependence often invokes, either explicitly or implicitly, limitations that are external to the mutualistic interaction itself \autocite{hale2021ecological, johnson2013competition}. However, it has been shown that intraspecific competition for partner resources can stabilize populations in models of two-species interactions \autocite{johnson2013competition}, explicitly linking dynamic stability to competitive interactions among mutualistic partners. Competition between and among mutualistic partners may play a central role in their ecological and evolutionary persistence \autocite{jones2012fundamental}.

Symbioses are a form of mutualism in which partners are in persistent contact \autocite{douglas2010symbiotic, douglas2015special}. In many symbioses, a larger host maintains an associated population of smaller symbionts \autocite{douglas2010symbiotic, douglas2015special}. It has been increasingly recognized that the nature of a given symbiosis can be context dependent, potentially shifting from mutualism to parasitism depending on external and internal factors \autocite{douglas2010symbiotic, douglas2015special}. In particular, the biomass ratio of the symbiotic partners can strongly determine the net cost and benefit of the interaction \autocite{becks2025emergent, cunning2014not}. The ecological stability and persistence of symbioses may thus depend on the population dynamics of symbionts with respect to an individual host. Further, the mutualistic outcome may depend on the mechanisms that regulate population dynamics of the symbionts including intraspecific (symbiont-symbiont) and interspecific (host-symbiont) competition for limiting resources \autocite{becks2025emergent}. However, it is unclear to what degree mechanisms such as intra- and interspecific competition shape mutualistic outcomes within symbioses \autocite{becks2025emergent}.

Many species within Cnidaria participate in a nutritional symbiosis, known as photosymbiosis, with dinoflagellates of the family Symbiodiniaceae \autocite{kayal2018phylogenomics, lajeunesse2018systematic}. The algal symbionts are hosted intracellularly, where they transfer photosynthetically fixed carbon to the host and gain access to limiting nutrients, largely as waste products from host metabolism \autocite{yellowlees2008metabolic, davy2012cell}. This symbiosis is especially conspicuous in scleractinian corals, where symbiont-derived photosynthate fuels host construction of calcium-carbonate skeletons that form the structural foundation of coral reefs \autocite{davy2012cell}.

It has been hypothesized that cnidarians, including scleractinian corals, maintain a stable population of algal symbionts by restricting the availability of nutrients critical to symbiont growth; specifically inorganic nitrogen in the form of ammonium \autocite{falkowski1993population, cui2019host, radecker2023coupled, xiang2020symbiont}. There is increasing evidence that this is achieved via host-mediated (re-)assimilation of ammonium fueled by symbiont-derived carbon \autocite{cui2019host, radecker2023coupled, xiang2020symbiont}. Specifically, symbiont-derived photosynthate fuels host amino-acid synthesis via the glutamine synthetase/glutamate synthase (GS–GOGAT) pathway \autocite{cui2019host, radecker2023coupled, xiang2020symbiont}. 

By fueling host assimilation of ammonium, symbionts may effectively limit their own density via facilitation of their competitor. The translocation of photosynthate to the host potentially creates a negative feedback loop wherein an increase in symbiont density leads to an increase in host assimilation of ammonium, thereby limiting further symbiont growth \autocite{cui2019host, radecker2023coupled}. Symbiont density may thus be regulated by a combination of intraspecific (symbiont-symbiont) and interspecific (host-symbiont) competition for ammonium. 

It has been proposed that this simple feedback loop could underlie the repeated evolution of photosymbiosis across a phylogenetically diverse range of taxa \autocite{cui2023carbon}. However, the sufficiency and dynamical consequences of this mechanism have not been well characterized. While some models have highlighted the importance of nitrogen limitation to the dynamic stability of the interaction \autocite{cunning2017dynamic, muller2009dynamic}, they included other limiting mechanisms, potentially obscuring the role of host-mediated ammonium assimilation.

Here, we ask if host-symbiont competition for ammonium is sufficient to yield dynamically stable symbiosis and, if so, whether the stabilizing mechanisms depend on the relative dominance of intra- versus interspecific competition. We consider a simple model of symbiont population regulation driven by a combination of intraspecific (symbiont-symbiont) and interspecific (host-symbiont) competition for ammonium \autocite{cui2019host, radecker2023coupled, radecker2021heat}. To characterize the dynamical stability of the model, we take the approach of generalized modeling \autocite{massing2022generalized, yeakel2011generalized, gross2006generalized}. Using generalized modeling, it is possible to characterize the local stability of steady states without specifying explicit functional forms for the processes within the model. This allows us to elucidate the key stabilizing feedback loops based only on the architecture of the interaction. We find that host-symbiont competition is not only sufficient to stabilize the interaction, but leads to a transition in the system wherein dynamic stability requires symbiont facilitation of host nitrogen assimilation via the translocation of photosynthate.

\section{Methods}

\subsection{A generalized model of host-symbiont nutrient competition}

\begin{figure}[h]
\centering
\includegraphics[width=0.4\textwidth]{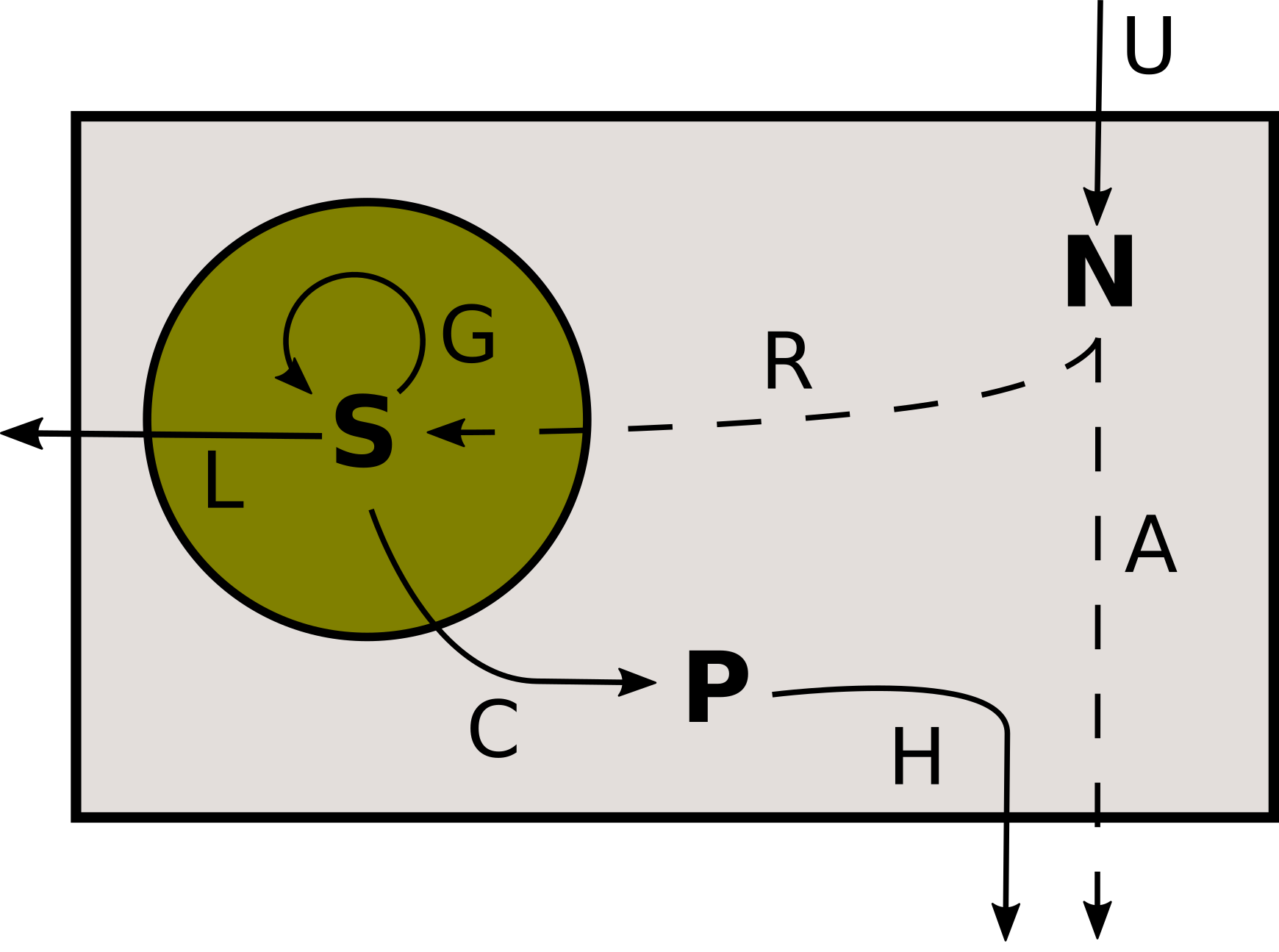}
\caption{Diagram of the model of host-symbiont nitrogen competition. Bold letters represent state variables while uppercase letters represent process functions. Nitrogen (N) enters the host where it is either assimilated (A) by the host or taken up (R) by the symbionts (S) for growth and reproduction. Photosynthate (P) is produced and translocated (C) by the symbionts where it is used (H) by the host to facilitate nitrogen assimilation.}
\label{fig:model-diagram}
\end{figure}

We consider the following simple model of coral-algal nutrient competition diagrammed in  Figure~\ref{fig:model-diagram}. Dissolved inorganic nitrogen in the form of ammonium (hereafter nitrogen) enters the system where it is either assimilated by the host or used by the symbionts for growth and reproduction. The symbionts constantly produce photosynthate, of which any excess is translocated to the host. The translocated photosynthate is then used by the host to facilitate nitrogen assimilation.

We can write the model as

\begin{align}
    \dot N &= U(N) - A(N, P) - R(N, S) \label{eqn:N} \\ 
    \dot S &= G(S, N) - L(S) \label{eqn:S} \\
    \dot P &= C(S, N) - H(N, P), \label{eqn:P}
\end{align}

where $N$ is the intracellular concentration of nitrogen, $S$ is the symbiont population density, and $P$ is the intracellular concentration of photosynthate. The dot above each variable denotes their derivative with respect to time. Nitrogen enters the system through an uptake function $U$ and is either assimilated by the host $A$ or utilized $R$ by the symbionts for growth and reproduction. Symbiont population density increases due to symbiont reproduction $G$ and decreases due to symbiont loss $L$ (which could represent expulsion from the host). Photosynthate is translocated $C$ by the symbionts and then used $H$ by the host to facilitate nitrogen assimilation.

Because the model focuses on the short-term dynamic regulation of the symbiont population in a healthy symbiotic state, we omit and/or condense some processes. First, we do not consider host growth as a dynamic variable. We instead assume the host has reached a steady state size at which point nitrogen assimilation fuels maintenance and metabolic turnover. Second, we omit nitrogen recycling as a dynamic input of nitrogen into the system. Host-assimilated nitrogen enters a slow-turnover pool where it is recycled back into the system on the order of months via host catabolism of metabolic products \autocite{tanaka2018stoichiometry}. However, in the healthy symbiotic state, the net flux of nitrogen into this pool is expected to be positive, representing a state of net positive host anabolism \autocite{radecker2021heat, cui2019host}. While sustained carbon limitation can induce a shift to net host catabolism \autocite{radecker2021heat}, the host can likely buffer short-term decreases in labile carbon to maintain a state of net anabolism in the healthy symbiotic state \autocite{radecker2018using}. We therefore expect the input of nitrogen via host catabolism to be weakly coupled to short-term fluctuations in labile carbon on the timescales considered here.

Now that we have specified the structure of the model---the state variables, the processes by which they change, and their relationship to one another---we can proceed to analyze the model to gain insights about the stability of the system.

\subsection{Model normalization and calculation of the Jacobian}

If we were to take the typical dynamical modeling approach, the next step would be to define the specific functional forms of the processes and choose biologically reasonable values for any parameters they might contain. We would then solve for the values of $N = N^*$, $S = S^*$, and $P = P^*$ that yield steady states where the rate of change in all state variables $\dot N = \dot S = \dot P = 0$. Finally, we would calculate the Jacobian to evaluate the local stability of the steady states.

Here we take the approach of generalized modeling to characterize the stability of the system \autocite{massing2022generalized, gross2006generalized, yeakel2011generalized}. Using generalized modeling, we do not specify functional forms for the processes. Rather, we formally normalize the model with respect to unknown, positive steady states. Normalizing the system in this manner allows us to parameterize the Jacobian in terms of normalized rates and elasticities (i.e. logarithmic derivatives of the unspecified process functions in the steady state). Elasticities capture the non-linearity in the response of each process to a small perturbation in the steady state. Because elasticities can be defined without reference to a specific functional form, we can characterize the local stability of any system that conforms to the structure in Eqs.~\eqref{eqn:N}--\eqref{eqn:P}.

To normalize the model, we begin by defining for each variable $X$

\begin{equation}
    x = \frac{X}{X^*}
\end{equation}

and for each process function $F(X)$

\begin{equation}
    f(x) = \frac{F(X)}{F^*} = \frac{F(xX^*)}{F^*}
\end{equation}

where $X^*$ denotes the value of $X$ in the steady state and $F^* = F(X^*)$ denotes the process function evaluated in the steady state.

Normalizing the model in Eqs.~\eqref{eqn:N}--\eqref{eqn:P} with respect to the steady states $[N^*, S^*, P^*]$ yields

\begin{align}
    \dot n &= \frac{U^*}{N^*} u(n) - \frac{A^*}{N^*} a(n, p) - \frac{R^*}{N^*} r(n, s) \label{eqn:norm_N} \\
    \dot s &= \frac{G^*}{S^*} g(s, n) - \frac{L^*}{S^*} l(s) \label{eqn:norm_S} \\
    \dot p &= \frac{C^*}{P^*} c(s, n) - \frac{H^*}{P^*} h(n, p) \label{eqn:norm_P}
\end{align}

where lowercase letters denote normalized process functions and state variables.

The benefit of normalizing the model in this fashion is that even though we did not specify functional forms for any of the processes, we now know the location of the steady states in the normalized system. Because each normalized state variable is defined as $x = X/X^*$, the steady state value of each variable $x^* = X^*/X^*$ equals 1. Additionally, because $f(x^*) = F^*/F^*$, all normalized process functions equal 1 in the steady state.

The prefactors of the normalized process functions in Eqs.~\eqref{eqn:norm_N}--\eqref{eqn:norm_P} denote the per-capita (or per-unit) rate of each process in the steady state. Because the rate of gain and loss must balance in the steady state, we can define the following scale parameters that represent the turnover rate for each state variable

\begin{align}
    \alpha_{\mathrm{n}} &= \frac{U^*}{N^*} = \frac{A^*}{N^*} + \frac{R^*}{N^*} \label{eqn:alpha_n} \\
    \alpha_{\mathrm{s}} &= \frac{G^*}{S^*} = \frac{L^*}{S^*} \label{eqn:alpha_s} \\
    \alpha_{\mathrm{p}} &= \frac{C^*}{P^*} = \frac{H^*}{P^*} \label{eqn:alpha_p}.
\end{align}

Nitrogen is lost through two different processes (Eq.~\ref{eqn:alpha_n}); assimilation by the host ($A$) and assimilation by the symbionts ($R$). We define the branching parameters $\beta$ and its complement $\bar \beta = 1 - \beta$, which represent the proportion of nitrogen lost through each process in the steady state

\begin{align}
    \beta &= \frac{1}{\alpha_{\mathrm{n}}} \frac{A^*}{N*} = \frac{A^*}{A^* + R^*} \\
    \bar \beta &= \frac{1}{\alpha_{\mathrm{n}}} \frac{R^*}{N*} = \frac{R^*}{A^* + R^*}.
\end{align}

The branching parameter $\beta$ represents the relative dominance of intraspecific (symbiont-symbiont) versus interspecific (host-symbiont) competition for nitrogen. At low values of $\beta$, the symbionts assimilate the majority of internal nitrogen in which case symbiont-symbiont competition for nitrogen is dominant. Increasing the proportion of nitrogen assimilated by the host, $\beta$, represents an increase in host-symbiont competition for internal nitrogen.

Substituting the scale and branching parameters into Eqs.~\eqref{eqn:norm_N}--\eqref{eqn:norm_P} allows us to write the final generalized model as

\begin{align}
    \dot n &= \alpha_{\mathrm{n}}(u(n) - \beta a(n, p) - \bar \beta r(n, s)) \label{eqn:n} \\
    \dot s &= \alpha_{\mathrm{s}} (g(s, n) - l(s)) \label{eqn:s} \\
    \dot p &= \alpha_{\mathrm{p}} (c(s, n) - h(n, p)) \label{eqn:p}.
\end{align}

Now that we have defined the generalized model, we can analyze the stability of the steady states in the normalized system. We begin by calculating the Jacobian matrix

\begin{equation} \label{eqn:Jacobian}
        \mathbf{J} = \begin{pmatrix}
        \alpha_{\mathrm{n}} (u_n - \beta a_n - \bar \beta r_n) & -\alpha_{\mathrm{n}} \bar \beta r_s & - \alpha_{\mathrm{n}} \beta a_p \\
        \alpha_{\mathrm{s}} g_n & \alpha_{\mathrm{s}}(g_s - l_s) & 0 \\
        \alpha_{\mathrm{p}}(c_n - h_n) & \alpha_{\mathrm{p}} c_s & -\alpha_{\mathrm{p}} h_p
    \end{pmatrix}
\end{equation}

where each entry $J_{ij}$ is the derivative of the $i$-th differential equation with respect to the $j$-th variable in the steady state

\begin{equation}
    \mathbf{J}_{ij} = \frac{\partial \dot x_i}{\partial x_j}\bigg\rvert_{*}.
\end{equation}
  
Each entry of the Jacobian describes how the time derivative of each state variable reacts to a given perturbation. For example, the entry $J_{11}$ captures the magnitude and direction of the rate of change of $\dot n$ following a small perturbation to the amount of nitrogen in the system in the steady state. 

The value of a given entry in the Jacobian depends upon the response of each of the relevant process functions to a small perturbation in the steady state. For example, in entry $J_{11}$, the parameters $u_n$, $a_n$, and $r_n$ capture the magnitude and direction of the response of the process functions  $u(n)$, $a(n)$, and $r(n)$ to a small change in nitrogen. Such parameters are known as elasticities. 

Elasticities are defined as logarithmic derivatives of the non-normalized process functions evaluated in the steady state. For a process function $F(X)$, the elasticity $f_x$ is defined as

\begin{equation}
    f_x = \frac{\partial \ln F(X)}{\partial \ln X} \bigg\rvert_{*}.
\end{equation}

Elasticities capture the degree of non-linearity in the response of a process function to a given perturbation in the steady state. Because we did not specify the process functions, the elasticities are treated as unknown parameters of the system.

The elasticities can be related in a simple and convenient way to the potential functional forms of the processes. For example, if the uptake process is a quadratic function of nitrogen such that $U(N) = bN^2$, then normalization will yield $u_n = 2$. Similarly, if the uptake function is an inverse function of nitrogen such that $U(N) = b / N$, then normalization will yield $u_n = -1$. In general, for any power law $F(X) = bX^k$, normalization will yield $f_x = k$.

The steady states of a dynamical system are locally stable if the real parts of all eigenvalues of its Jacobian are negative. We can therefore evaluate the stability of the symbiotic interaction by calculating the eigenvalues of the Jacobian. Because the parameters of the Jacobian represent biologically interpretable processes, we can now characterize how these processes affect the stability of the symbiosis. 

\subsection{Numerical stability analysis}

To explore how the processes in our model affect stability across a range of biologically plausible functional responses, we used an ensemble approach as described in \textcite{massing2022generalized, yeakel2011generalized}. For each parameter, we generated $M = 10^7$ random samples from a uniform distribution with ranges specified in Table~\ref{tab:par}. For each of the $M$ parameter sets, we populated the Jacobian and calculated its eigenvalues. We defined

\begin{equation}
s_m =
\begin{cases}
1, & \operatorname{Re}(\lambda_0) < 0 \\
0, & \text{otherwise}
\end{cases}
\end{equation}

such that $s_m$ equals 1 if the $m$-th parameter set is stable (i.e. if the real part of the Jacobian's leading eigenvalue is negative) and equals 0 otherwise.

Because the proportion of nitrogen assimilated by the host, $\beta$, represents the relative dominance of symbiont-symbiont versus host-symbiont competition for internal nitrogen, shifts in $\beta$ may result in concomitant shifts in the identity of stabilizing processes. To explore the relationship between host assimilation and stability, we treated the proportion of nitrogen assimilated by the host, $\beta$, as the organizing axis for the subsequent correlation analysis. By considering stability as a function of $\beta$, we can understand how the relative dominance of intraspecific (symbiont-symbiont) versus interspecific (host-symbiont) competition for nitrogen affects the stability of the interaction.

We binned the $M$ parameter sets into $B = 20$ bins based on their sampled values of $\beta$. Bins were of equal width $0.05$. Within each bin, we calculated the Pearson correlation between the $m$-th realization of each parameter (excluding $\beta$) and the stability of the $m$-th parameter set $s_m$. Large positive values of the Pearson correlation indicate that larger values of the parameter are stabilizing while large negative values of the Pearson correlation indicate that smaller values of the parameter are stabilizing.

\subsection{Bifurcation analysis}

To understand in more detail how individual processes affect the stability of the system, we analytically searched for bifurcation points in the parameter space. A bifurcation is a qualitative change in the long-term dynamics of the system, such as a loss of stability or a change in the number or types of long-term behavior.

Bifurcations of steady states typically occur in two ways: fold (saddle-node) bifurcations and Hopf bifurcations \autocite{guckenheimer2013nonlinear}. Fold bifurcations represent the destruction of steady states and can represent runaway growth or the collapse of interactions \autocite{massing2025generalized}. Hopf bifurcations indicate the onset of oscillations which can represent runaway oscillatory behavior or the onset of stable limit cycles \autocite{massing2022generalized, yeakel2011generalized}.

Fold bifurcations occur when a single, real eigenvalue crosses the imaginary axis to become positive; i.e. when the Jacobian has a zero eigenvalue \autocite{guckenheimer2013nonlinear}. Because the determinant of a matrix is the product of its eigenvalues, we set the determinant of the Jacobian equal to 0 as a test function for fold bifurcations \autocite{massing2022generalized, stiefs2008computation}. 

Hopf bifurcations occur when a complex-conjugate pair of eigenvalues simultaneously cross the imaginary axis such that the real parts become positive \autocite{guckenheimer2013nonlinear}. We used the method of resultants \autocite{gross2004analytical, guckenheimer1997computing} to derive a test function for Hopf bifurcations. 

We solved the test functions for fold and Hopf bifurcations symbolically to plot bifurcation points for a subset of parameters identified as important by the numerical stability analysis. To understand the relationship between stability and the relative dominance of intraspecific (symbiont-symbiont) versus interspecific (host-symbiont) competition, we generated bifurcation points for each focal parameter as a function of the proportion of nitrogen assimilated by the host, $\beta$. Non-focal parameters were fixed to biologically plausible values representing a healthy symbiotic state as listed in Table~\ref{tab:par}. We restricted ourselves to the case where symbiont density-dependent growth is stronger than density-dependent loss ($g_s > l_s$) in the steady state, which allows us to understand the stability of the system when nitrogen limitation is the only mechanism available to control symbiont population density.

\begin{table}
    \centering
    \caption{Parameters of the Jacobian, their range of values for the numerical stability analysis, their fixed values for the bifurcation analysis, their definition, and their fixed-value interpretation. The symbol $\delta_X$ denotes a small, positive perturbation in variable $X$ in the steady state.}
    \label{tab:par}
    \small
    \begin{tabular}{>{\raggedright\arraybackslash}p{0.15\linewidth}>{\raggedright\arraybackslash}p{0.08\linewidth}>{\raggedright\arraybackslash}p{0.08\linewidth}>{\raggedright\arraybackslash}p{0.3\linewidth}>{\raggedright\arraybackslash}p{0.3\linewidth}}\toprule
         \textbf{Parameter} &  \textbf{Range}&  \textbf{Value}&  \textbf{Definition}& \textbf{Value Interpretation}\\\midrule
         $\alpha_{\mathrm{n}}$         &  [1,10]         &  10             &  Nitrogen turnover rate                                       & Fast relative to symbiont turnover\\
         $\beta$          &  [0,1]          &  [0,1]&  Proportion of nitrogen assimilated by the host                          & Dominance of intra- vs interspecific competition\\
         $u_n$              &  [-2,2]         &  -0.5             &  Response of nitrogen uptake to $\delta_N$& Decreases sub-linearly (moderately regulated)\\
         $a_n$              &  [0,2]          &  1              &  Response of host nitrogen assimilation to $\delta_N$& Increases linearly\\
         $a_p$              &  [0,2]          &  1              &  Response of host nitrogen assimilation to $\delta_P$& Increases linearly\\
         $r_n$              &  [0,2]          &  1              &  Response of symbiont nitrogen assimilation to $\delta_N$& Increases linearly\\
         $r_s$              &  [0,2]          &  1              &  Response of symbiont nitrogen assimilation to $\delta_S$& Increases linearly\\
         $\alpha_{\mathrm{s}}$         &  [1,10]         &  1              &  Symbiont turnover rate                                       & Slow relative to N and P turnover\\
 $g_s$              & [0,2]          & 1              & Response of symbiont growth to $\delta_S$&Increases linearly\\
 $l_s$              & [0,2]          & 0.2            & Response of symbiont loss to $\delta_S$&Increases sub-linearly (saturated expulsion)\\
 $g_n$              & [0,2]          & 1.5              & Response of symbiont growth to $\delta_N$&Increases super-linearly (symbionts are nitrogen limited)\\
 $\alpha_{\mathrm{p}}$         & [1,10]         & 10              & Photosynthate turnover rate                                  &Fast relative to symbiont turnover\\
 $c_s$              & [-2,2]         & 1              & Response of photosynthate translocation to $\delta_S$&Increases linearly\\
 $c_n$              & [-2,0]        & -1             & Response of photosynthate translocation to $\delta_N$&Decreases linearly\\
 $h_n$              & [0,2]          & 1              & Response of host photosynthate use to $\delta_N$&Increases linearly\\
 $h_p$              & [0,2]          & 2              & Response of host photosynthate use to $\delta_P$&Increases quadratically (multiple metabolic sinks)\\ \bottomrule
    \end{tabular}
\end{table}

\FloatBarrier
\section{Results}

\subsection{Nitrogen limitation is necessary for dynamic stability.}

\begin{figure}[h!]
\centering
\includegraphics[width = \textwidth]{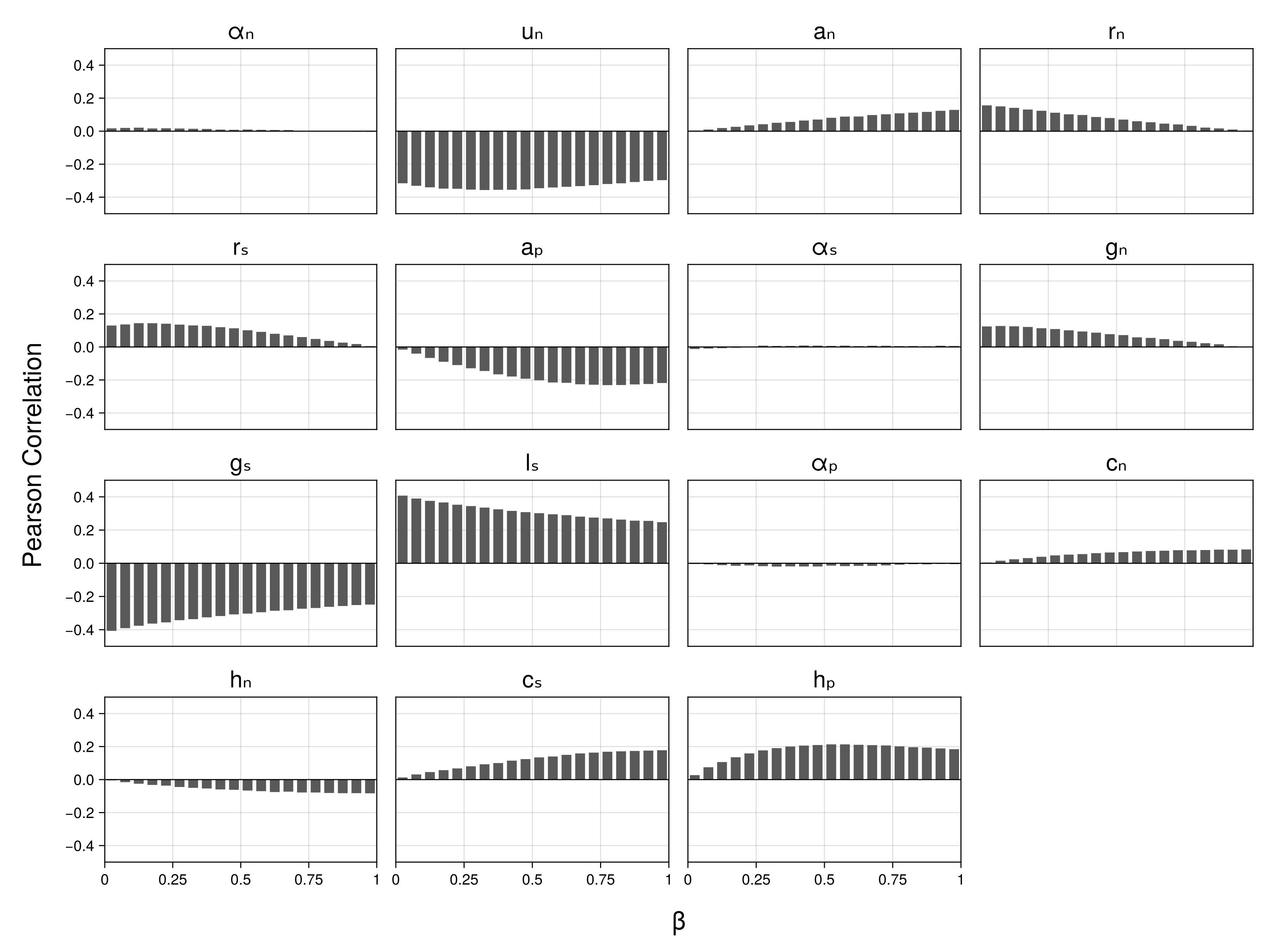}
\caption{Correlation of rate and elasticity parameters with stability. Correlations are binned by proportion of nitrogen assimilated by the host, $\beta$. Positive correlations indicate that higher values of the parameter promote stability while negative correlations indicate that lower values of the parameter promote stability. Note that for parameters restricted to a positive range, a negative correlation indicates that a weaker positive response is stabilizing relative to a stronger positive response. Ranges for all parameters are listed in Table~\ref{tab:par}.}
\label{fig:beta-histogram}
\end{figure}

The numerical stability analysis identified three processes that are important in stabilizing the system across the entire range of proportion of nitrogen assimilated by the host $\beta$ (Fig.~\ref{fig:beta-histogram}):
\begin{enumerate}
    \item Nitrogen uptake with respect to a change in internal nitrogen availability ($u_n$).
    \item Symbiont growth with respect to a change in symbiont density ($g_s$).
    \item Symbiont loss with respect to a change in symbiont density ($l_s$).
\end{enumerate}

The strongly negative correlation of $u_n$ with stability indicates that given a small increase in internal nitrogen availability, a decrease in the uptake of external nitrogen promotes stability. The parameter $u_n$ shows a strong correlation with stability across the entire range of proportion of nitrogen assimilated by the host, indicating that nitrogen limitation is key to stabilizing the system whether symbiont-symbiont or host-symbiont competition is dominant.

The negative correlation of $g_s$ and the positive correlation of $l_s$ with stability indicate that given a small increase in symbiont density, a stronger response in symbiont loss relative to symbiont growth promotes stability. In other words, it is stabilizing if symbiont growth is limited by an additional factor independent of nitrogen availability (e.g. space limitation within the host or an increase in symbiont expulsion by the host). In the bifurcation analysis below, we consider stability when nitrogen limitation is the only factor limiting symbiont growth (i.e. when $g_s > l_s$).

\subsection{Symbiont growth and subsequent nitrogen consumption promote stability when interspecific competition is weak.}

The numerical stability analysis also identified processes that are important in stabilizing the system depending on the proportion of nitrogen assimilated by the host, $\beta$.

When the proportion of nitrogen assimilated by the host is low ($\beta < 0.5$), the following processes related to symbiont growth and nitrogen uptake are relatively important in stabilizing the system (Fig.~\ref{fig:beta-histogram}):
\begin{enumerate}
    \item Symbiont growth with respect to a change in internal nitrogen availability ($g_n$).
    \item Symbiont nitrogen uptake with respect to a change in internal nitrogen availability ($r_n$).
    \item Symbiont nitrogen uptake with respect to a change in symbiont density ($r_s$).
\end{enumerate}

The positive correlation of $g_n$ and $r_n$ with stability indicates that given an increase in nitrogen, an increase in both symbiont growth and symbiont assimilation of nitrogen promotes stability. The positive correlation of $r_s$ with stability indicates that given an increase in symbiont density, an increase in symbiont nitrogen uptake promotes stability. Taken together, these results suggest that when interspecific competition is weak, stability is maintained by nitrogen limitation achieved through a combination of symbiont growth and subsequent symbiont use of nitrogen.

\subsection{Symbiont facilitation of host nitrogen assimilation promotes stability when interspecific competition is strong.}

As the proportion of nitrogen assimilated by the host increases, the following processes related to host assimilation of nitrogen and symbiont translocation of photosynthate become increasingly important in stabilizing the system (Fig.~\ref{fig:beta-histogram}):
\begin{enumerate}
    \item Host nitrogen assimilation with respect to a change in internal nitrogen availability ($a_n$).
    \item Host nitrogen assimilation with respect to a change in photosynthate availability ($a_p$).
    \item Translocation of photosynthate with respect to a change in symbiont density ($c_s$).
    \item Host photosynthate use with respect to a change in photosynthate availability ($h_p$).
\end{enumerate}

The positive correlation of $a_n$ with stability indicates that given an increase in nitrogen, an increase in host assimilation of nitrogen promotes stability. The negative correlation of $a_p$ with stability indicates that given an increase in photosynthate, a weaker increase in host assimilation of nitrogen promotes stability  ($a_p$ is restricted to a range of positive values so a negative correlation indicates that a weaker positive response is stabilizing).

The positive correlation of $c_s$ indicates that given an increase in symbiont density, an increase in translocation of photosynthate promotes stability. Finally, the positive correlation of $h_p$ with stability indicates that given an increase in photosynthate, an increase in host use of photosynthate promotes stability. This likely indicates that a buildup of photosynthate through inefficient host use can destabilize the system by decoupling host assimilation from symbiont density---a hypothesis we further explore in the bifurcation analysis below. Taken together, these results suggest that as interspecific competition increases, stability is maintained by symbiont facilitation of host nitrogen assimilation via the translocation of photosynthate.

\subsection{Nitrogen limitation is sufficient for dynamic stability when symbiont density is not self-limiting.}

\begin{figure}[H]
\centering
\includegraphics[width = \textwidth]{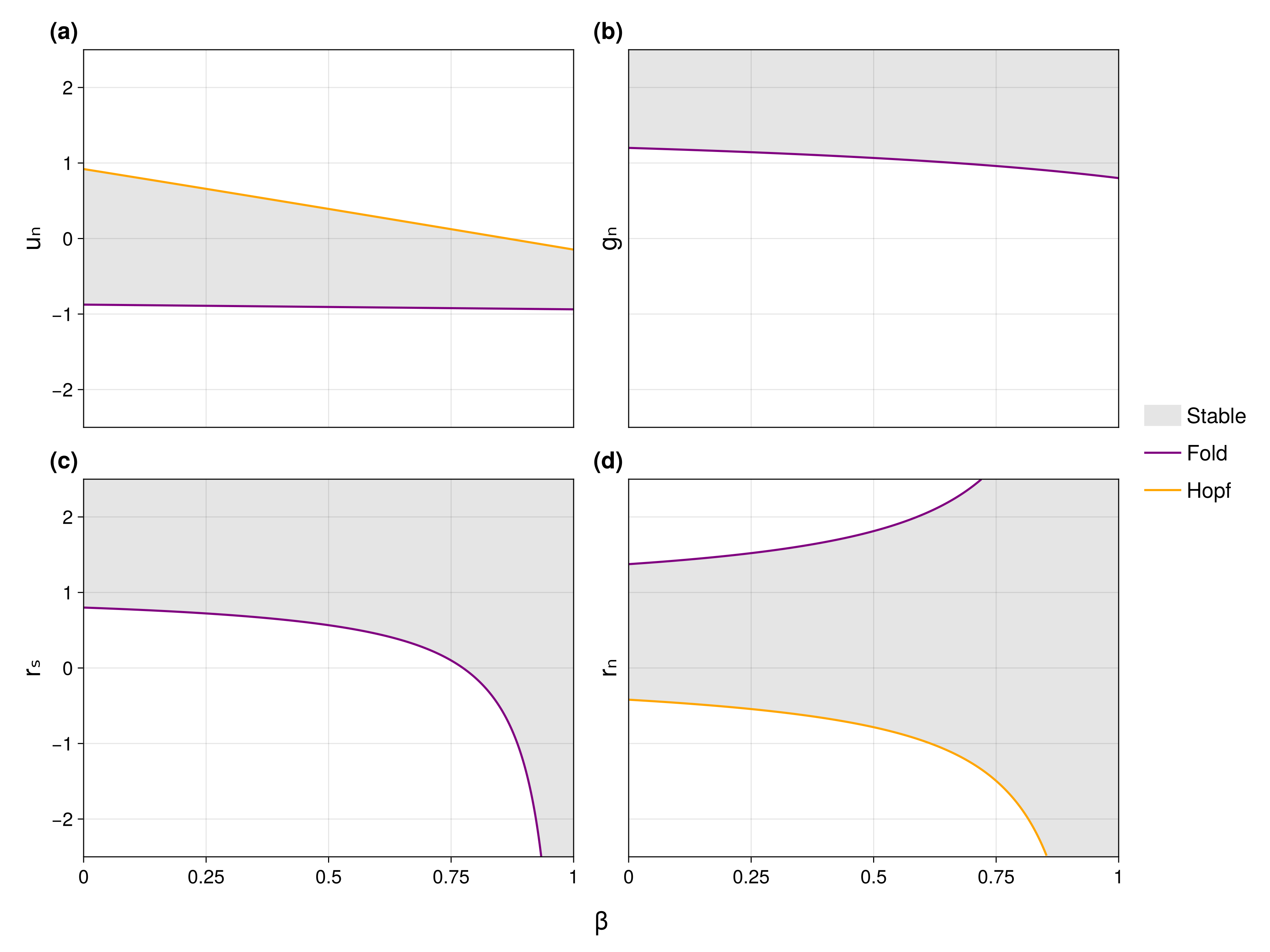}
\caption{Bifurcation plots for processes related to nitrogen uptake, symbiont growth and symbiont assimilation of nitrogen. Orange lines denote Hopf bifurcations, purple lines denote fold bifurcations, and gray shading denotes region of stability. (a) Given an increase in internal nitrogen availability, a sub-linear decrease in nitrogen uptake promotes stability across the entire range of $\beta$. (b) Super-linear symbiont growth with respect to nitrogen and (c) nearly linear nitrogen use with respect to symbiont density are necessary to maintain stability. (d) Symbiont nitrogen assimilation with respect to a change in nitrogen does not strongly affect stability, likely due to increased assimilation driven by strong symbiont growth.}
\label{fig:bifurcation-uptake-growth}
\end{figure}

While the numerical stability analysis is an efficient way to identify the nature and identity of processes that are important in stabilizing the system, the bifurcation analysis provides a more detailed understanding of how and where stability is lost in the parameter space with respect to individual processes. Additionally, by restricting ourselves to the case where symbiont growth is not directly limited by their own density ($l_s < g_s$; Table~\ref{tab:par}), we can explore how the system is stabilized when nitrogen limitation is the only potential regulatory mechanism. Under this restriction, we find that the interaction can still be dynamically stable. A combination of limited nitrogen uptake by the host and strong growth and subsequent nitrogen use by the symbionts promotes stability through nitrogen limitation (Fig.~\ref{fig:bifurcation-uptake-growth}).

The bifurcation diagram for $u_n$ (Fig.~\ref{fig:bifurcation-uptake-growth}a) shows that given an increase in internally available nitrogen in the steady state, the uptake of nitrogen must respond, at most, sub-linearly to avoid crossing a Hopf bifurcation for $\beta < 0.85$. This suggests that if the system responds to an increase in nitrogen by a sufficiently strong increase in uptake, stability is lost through oscillations, likely driven by overshoot and subsequent crashes of the symbiont population in response to the initial strong increase and subsequent depletion of nitrogen. Stability is possible across the entire range of $\beta$ when uptake decreases sub-linearly given an increase in internally available nitrogen, indicating that moderately regulated uptake promotes stability. 

The bifurcation diagram for $g_n$ (Fig.~\ref{fig:bifurcation-uptake-growth}b) shows that given an increase in nitrogen in the steady state, symbiont growth must respond strongly (super-linearly)  for $\beta < 0.65$ to avoid crossing a fold bifurcation. Additionally, the bifurcation diagram for $r_s$ (Fig.~\ref{fig:bifurcation-uptake-growth}c) indicates that given an increase in symbiont density, symbiont nitrogen use should increase at least nearly linearly for most of the range of $\beta$ to avoid crossing a fold bifurcation. Taken together, this indicates that strong symbiont growth and subsequent nitrogen use are necessary to avoid system collapse, especially when the host assimilates less than 65\% of the nitrogen. Additionally, a super-linear response in symbiont growth with respect to nitrogen suggests that the population is well below its growth potential with respect to nitrogen, indicating that a nitrogen-limited symbiont population is necessary for stability. 

Finally, the bifurcation diagram for $r_n$ (Fig.~\ref{fig:bifurcation-uptake-growth}d) shows that given an increase in nitrogen, symbiont use of nitrogen can take on a relatively large range of responses (including an elasticity of $0$). While seemingly counterintuitive, this  suggests that as long as nitrogen uptake responds sufficiently negatively ($u_n = -0.5$; Table~\ref{tab:par}), and symbiont growth responds sufficiently positively ($g_n$ = 1.5; Table~\ref{tab:par}), nitrogen limitation is maintained by an increase in symbionts ($r_s$ = 1, Table~\ref{tab:par}), rendering $r_n$ less critical for dynamic stability.

\subsection{Symbiont facilitation of host nitrogen assimilation becomes necessary for dynamic stability as interspecific competition increases.}

\begin{figure}[H]
\centering
\includegraphics[width = \textwidth]{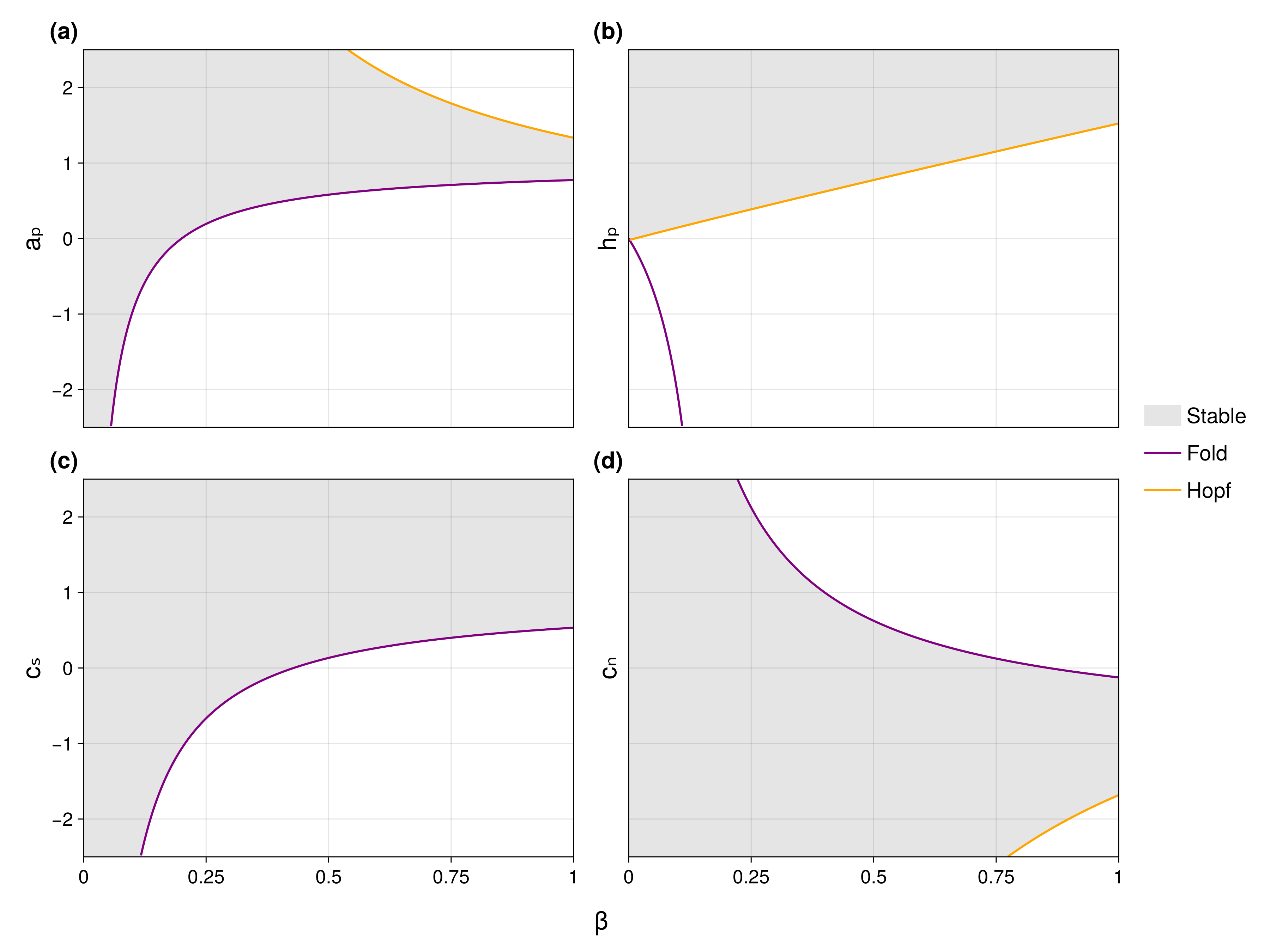}
\caption{Bifurcation plots for processes associated with host nitrogen assimilation, host use of photosynthate, and symbiont translocation of photosynthate. Orange lines denote Hopf bifurcations, purple lines denote fold bifurcations, and gray shading denotes region of stability. (a) Host assimilation of nitrogen must respond positively to an increase in photosynthate when interspecific competition is sufficiently strong ($\beta > 0.2$). (b) Host use of photosynthate must respond positively to an increase in photosynthate availability. (c) Translocation of photosynthate must respond positively to an increase in symbiont density when interspecific competition is sufficiently strong ($\beta > 0.43$). (d) Translocation of photosynthate must respond negatively to an increase in nitrogen when interspecific competition is high ($\beta > 0.85$).}
\label{fig:bifurcation-assimilation-photosynthate}
\end{figure}

The bifurcation analysis also identifies a transition where dynamic stability requires symbiont facilitation of host nitrogen assimilation. As interspecific (host-symbiont) competition for nitrogen increases, host assimilation of nitrogen must be coupled to changes in symbiont density via the translocation of photosynthate for the interaction to be dynamically stable (Fig.~\ref{fig:bifurcation-assimilation-photosynthate}).

The bifurcation diagram for $a_p$ (Fig.~\ref{fig:bifurcation-assimilation-photosynthate}a) shows that given an increase in photosynthate, the response of host nitrogen assimilation does not affect dynamic stability for $\beta < 0.2$. However, when $\beta = 0.2$, the fold bifurcation crosses $0$, indicating that host assimilation must respond positively to an increase in photosynthate for the system to be stable. This transition marks the point at which host assimilation of nitrogen must be coupled to the translocation of photosynthate to maintain stability. When the host assimilates the majority of internally available nitrogen ($\beta > 0.5$), a super-linear increase in assimilation risks loss of stability by crossing a Hopf bifurcation, which may represent oscillations driven by rapid depletion of available photosynthate followed by decreased assimilation and a consequent increase in nitrogen available for symbiont growth.

The bifurcation diagram for $h_p$ (Fig.~\ref{fig:bifurcation-assimilation-photosynthate}b) shows that given an increase in photosynthate, host use of photosynthate must respond positively to avoid crossing a Hopf bifurcation. Inefficient host use of photosynthate may cause it to accumulate, leading to oscillations as available photosynthate becomes increasingly decoupled from the current symbiont population density.

 The bifurcation diagram for $c_s$ (Fig.~\ref{fig:bifurcation-assimilation-photosynthate}c) shows that given an increase in symbiont density, the response of translocation does not affect dynamic stability when $\beta$ is sufficiently low ($\beta < 0.1)$. However, when the host assimilates a sufficient proportion of nitrogen ($\beta \approx 0.43$), translocation of photosynthate must respond positively to an increase in symbiont population density to avoid crossing a fold bifurcation. Crossing the fold bifurcation may represent runaway symbiont growth as an increase in population density does not yield enough photosynthate to limit nitrogen availability via host assimilation.

Finally, the bifurcation diagram for $c_n$ (Fig.~\ref{fig:bifurcation-assimilation-photosynthate}d) shows that given an increase in nitrogen, symbiont translocation should decrease to avoid crossing a fold bifurcation when $\beta$ is sufficiently high ($\beta \approx 0.85$). This may indicate that, given an increase in nitrogen, a positive response in translocation could lead to runaway nitrogen depletion by the host.

\section{Discussion}

Symbioses represent a balancing act between the costs and benefits of maintaining an associated population of symbionts \autocite{becks2025emergent, cunning2014not, douglas2010symbiotic}. Because symbiont density determines the net benefit of the interaction, the ecological stability of symbiosis depends on the mechanisms by which the symbiont population is regulated \autocite{becks2025emergent, cunning2014not}. The nature of these regulatory mechanisms likely depends on the type of symbiosis and resources exchanged. In plant-mycorrhizal interactions, previous work has highlighted the importance of active host sanctions to differentially punish and facilitate symbionts based on their degree of cooperation \autocite{douglas2010symbiotic}. In cnidarian-algal photosymbiosis, symbiont density is potentially regulated through a passive feedback loop wherein cooperation is enforced via host-symbiont competition for a single limiting resource \autocite{cui2019host, cui2023carbon, radecker2023coupled, xiang2020symbiont}. 

Here, we show that host-symbiont competition for inorganic nitrogen in the form of ammonium is sufficient to yield dynamically stable symbiosis. Moreover, nitrogen limitation is sufficient to stabilize symbiont density absent any other form of density dependence. Most importantly, we find that the identity of the processes that stabilize the interaction depends on the proportion of nitrogen assimilated by the host. When the proportion of nitrogen assimilated by the host is low, stability is maintained via symbiont growth and subsequent nitrogen use. As the proportion of nitrogen assimilated by the host increases, stability increasingly requires that symbiont density is coupled to host assimilation of nitrogen via the translocation of photosynthate.

When the proportion of nitrogen assimilated by the host is sufficiently low, both numerical stability (Fig.~\ref{fig:beta-histogram}) and bifurcation analyses (Fig.~\ref{fig:bifurcation-uptake-growth}) indicate that intraspecific (symbiont-symbiont) competition for nitrogen is sufficient to stabilize the interaction. In the steady state, local stability is maintained through strong symbiont growth given an increase in nitrogen and strong nitrogen use given an increase in symbiont density. Symbiont growth is limited largely by cell-cycle arrest, with up to 90\% of symbionts arrested in G1 phase, consistent with strong nutrient limitation \autocite{gorman2025stability}. Additionally, experimental addition of ammonium leads to rapid assimilation by symbionts \autocite{pernice2012single} and increases in symbiont density \autocite{radecker2023coupled}. Taken together, experimental evidence \autocite{gorman2025stability, pernice2012single, radecker2023coupled} and the results of this study suggest that nitrogen limitation is enforced through strong symbiont growth and subsequent nitrogen use.

As the proportion of nitrogen assimilated by the host increases, we find that dynamic stability increasingly requires that symbionts facilitate host assimilation of nitrogen through the translocation of photosynthate. When the proportion of nitrogen assimilated by the host is sufficiently high, host assimilation of nitrogen must be positively coupled to the translocation of photosynthate (Fig.~\ref{fig:bifurcation-assimilation-photosynthate}a). Additionally, symbiont density must be positively coupled to translocation of photosynthate (Fig.~\ref{fig:bifurcation-assimilation-photosynthate}c). These positive couplings indicate that high interspecific competition for internal nitrogen is maintained by symbiont facilitation of host nitrogen assimilation. The coupling between translocation of photosynthate and interspecific competition for nitrogen was demonstrated experimentally in \textit{Aiptasia} by observing that the addition of labile carbon increases host assimilation of nitrogen and decreases symbiont density \autocite{radecker2023coupled, cui2019host}. Notably, the results of the bifurcation analysis indicate that symbiont facilitation of interspecific competition is necessary for stability when the host assimilates $\approx40\%$ of the available nitrogen (Fig.~\ref{fig:bifurcation-assimilation-photosynthate}a and Fig.~\ref{fig:bifurcation-assimilation-photosynthate}c), consistent with empirical estimates of host assimilation \autocite{cui2023molecular} in \textit{Aiptasia}. 

Regardless of whether intra- or interspecific competition dominates, we find that stability depends upon how host uptake of external nitrogen responds to changes in the concentration of internal nitrogen. If there is an increase in the concentration of internal nitrogen, a sufficiently strong increase in host uptake of external nitrogen can destabilize the system (Fig.~\ref{fig:beta-histogram} and Fig.~\ref{fig:bifurcation-uptake-growth}a). When interspecific competition is low, the system can tolerate a weak increase in host uptake of external nitrogen in response to an increase in internal nitrogen, suggesting that strong symbiont growth and subsequent nitrogen depletion may be sufficient to maintain stability. However, as interspecific competition increases, host uptake of external nitrogen must be increasingly regulated to avoid instability, suggesting that an uncontrolled influx of external nitrogen can overwhelm the host's ability to limit internal nitrogen through assimilation. Consistent with this interpretation, eutrophication has been shown to interfere with nutrient cycling and increase susceptibility to environmental stress in scleractinian corals \autocite{radecker2021heat}.
 
Taken together, intra- and interspecific competition for nitrogen are sufficient to stabilize the interaction. In fact, when the proportion of internal nitrogen assimilated by the host is low, our results suggest that intraspecific competition alone for nitrogen is sufficient to regulate symbiont population density. Previous studies have demonstrated that intraspecific competition for mutualistic benefits can stabilize populations in two-species models of mutualism \autocite{johnson2013competition}. This begs the question: why, and when, is interspecific competition necessary to stabilize the interaction?

It is important to note that, while the proportion of nitrogen assimilated by the host ($\beta$) is a static parameter in the generalized model, it is a dynamic variable in the real interaction. In reality, low values of $\beta$ (i.e. weak interspecific competition) are expected to be a transient feature of the interaction \autocite{cui2023carbon, radecker2023coupled}. Consider that in a nitrogen-replete intracellular environment, symbionts will invest most photosynthetically fixed carbon into fueling their own growth, with correspondingly little photosynthate translocated to the host. Consequently, the host will have limited capacity to assimilate intracellular nitrogen. However, as the symbiont population increases, intracellular nitrogen will become increasingly limited for the symbionts. Symbiont population growth will therefore decrease with a concomitant increase in translocated photosynthate to the host. Host assimilation of intracellular nitrogen will consequently increase, further limiting nitrogen availability for the symbionts. Thus, intraspecific competition between symbionts for intracellular nitrogen will naturally drive a transition to increased interspecific competition via increased translocation of photosynthate. 

Considering the dynamic nature of $\beta$, the bifurcation diagrams are best viewed as maps that show how the identity and nature of stabilizing processes depend on the proportion of nitrogen assimilated by the host. We can therefore imagine the system moving naturally along the $\beta$ axis as increased translocation of photosynthate facilitates increased interspecific competition for nitrogen. The bifurcation diagrams thus demonstrate that the natural transition from intra- to interspecific competition necessitates the emergence of a positive coupling of symbiont density to host nitrogen assimilation via photosynthate translocation for dynamic stability (Fig.~\ref{fig:bifurcation-assimilation-photosynthate}a and Fig.~\ref{fig:bifurcation-assimilation-photosynthate}c).  

The dynamic transition from intra- to interspecific competition and the concordant necessity for a positive coupling of symbiont density to host nitrogen assimilation via photosynthate translocation hint at a connection between the ecological stability and evolutionary onset of the symbiosis. Because this positive coupling does not require the evolution of novel mechanisms to regulate the symbiosis, it has been proposed that it may underlie the repeated evolution of photosymbiosis across a phylogenetically diverse range of organisms \autocite{cui2023carbon}. Additionally, by-product interactions are thought to be an important stepping-stone in the evolutionary origin of some symbioses \autocite{douglas2010symbiotic, douglas2015special}. In the cnidarian-algal symbiosis, host waste-ammonia may have served as the initial by-product resource that facilitated the onset of the interaction \autocite{douglas2015special}. After initial infection and rapid growth of the algal symbionts, nitrogen limitation would result in increased translocation of unused photosynthate to the host, driving increased interspecific competition via a transition to symbiont facilitation of host nitrogen assimilation.

While our results demonstrate that host-symbiont competition for nitrogen is sufficient to stabilize the interaction, this does not mean that other forms of negative feedback or host regulation are absent or unimportant. Other limiting resources, such as phosphorus and inorganic carbon, are likely important in driving symbiont dynamics \autocite{cunning2017dynamic, muller2009dynamic}. Additionally, other mechanisms of regulation such as expulsion of symbionts by the host may be important depending on the stage of the algal infection and on environmental conditions \autocite{gorman2025stability, dimond2008symbiosis}. However, mechanisms such as host expulsion may represent a more costly alternative to nitrogen limitation through host assimilation. In this sense, interspecific competition for nitrogen may represent the most cost-efficient way to maintain symbiont density under healthy conditions.

It is also important to note that our model does not consider coral bleaching; i.e. the mass expulsion of symbionts by the host. In the context of our model, loss of stability represents the breakdown of nutrient cycling and subsequent loss of control of symbiont density. However, it has been shown that a breakdown of nutrient cycling may precede bleaching \autocite{radecker2021heat}. Thus, the loss of stability in our model could represent the initial loss of control of symbiont density preceding the more drastic and energetically costly mechanism of mass expulsion by the host. 

While the generalized model identifies the processes that are important in stabilizing the interaction, due to the normalization procedure, it cannot explicitly quantify changes in steady state values of the state variables nor simulate dynamic trajectories or outcomes. However, our results offer a starting point from which more fully specified models can be derived within the constraints identified here for dynamic stability. In fact, the lack of specific functional forms can be seen as a strength of the present approach. Photosymbiotic organisms are phylogenetically diverse and the particular functional relationships between different processes may be varied as well. By focusing on the qualitative reaction of the various processes, we have identified the core processes necessary to stabilize the interaction regardless of specific functional form. Additionally, the stabilizing processes identified here can serve as the foundation upon which more taxon-specific layers of complexity can be built. 

A central challenge in the study of symbioses, and mutualisms in general, has been to understand how system-specific details and general ecological mechanisms interact to shape their ecological and evolutionary persistence \autocite{douglas2010symbiotic, douglas2015special, bronstein2015study, sachs2006pathways}.  When seen as a consumer-resource interaction, the factors that stabilize symbioses are largely similar to the factors that stabilize more familiar competitive and consumer interactions \autocite{holland2015population, jones2012fundamental, johnson2013competition}. Here, we demonstrate how a combination of intra- and interspecific competition for a single resource is sufficient to stabilize cnidarian-algal photosymbiosis. Most importantly, we show that increased interspecific competition requires the emergence of a positive coupling of symbiont density to host nitrogen assimilation via photosynthate translocation for dynamic stability. This transition arises naturally from the dynamics of the interaction, reinforcing the possibility that this simple mechanism underlies the repeated independent evolution of photosymbiosis \autocite{cui2023carbon}. It is tempting to speculate that similar mechanisms may underlie the evolutionary onset and ecological persistence of other symbioses, and a future challenge will be to identify the extent to which symbiotic interactions are mediated by such simple ecological feedbacks.

\clearpage
\printbibliography

@article{massing2022generalized,
author={Massing, Jana C.  and Gross, Thilo },    
title={Generalized Structural Kinetic Modeling: A Survey and Guide},     
journal={Frontiers in Molecular Biosciences},     
volume={Volume 9 - 2022},
year={2022},
url={https://www.frontiersin.org/journals/molecular-biosciences/articles/10.3389/fmolb.2022.825052},
doi={10.3389/fmolb.2022.825052},
issn={2296-889X}}

@article{gross2004analytical,
  title={Analytical search for bifurcation surfaces in parameter space},
  author={Gross, Thilo and Feudel, Ulrike},
  journal={Physica D: Nonlinear Phenomena},
  volume={195},
  number={3-4},
  pages={292--302},
  year={2004},
  publisher={Elsevier}
}

@article{yeakel2011generalized,
  title={Generalized modeling of ecological population dynamics},
  author={Yeakel, Justin D and Stiefs, Dirk and Novak, Mark and Gross, Thilo},
  journal={Theoretical Ecology},
  volume={4},
  pages={179--194},
  year={2011},
  publisher={Springer}
}

@article{cui2019host,
  title={Host-dependent nitrogen recycling as a mechanism of symbiont control in Aiptasia},
  author={Cui, Guoxin and Liew, Yi Jin and Li, Yong and Kharbatia, Najeh and Zahran, Noura I and Emwas, Abdul-Hamid and Eguiluz, Victor M and Aranda, Manuel},
  journal={PLoS genetics},
  volume={15},
  number={6},
  pages={e1008189},
  year={2019},
  publisher={Public Library of Science San Francisco, CA USA}
}

@article{davy2012cell,
  title={Cell biology of cnidarian-dinoflagellate symbiosis},
  author={Davy, Simon K and Allemand, Denis and Weis, Virginia M},
  journal={Microbiology and Molecular Biology Reviews},
  volume={76},
  number={2},
  pages={229--261},
  year={2012},
  publisher={American Society for Microbiology 1752 N St., NW, Washington, DC}
}

@article{radecker2023coupled,
  title={Coupled carbon and nitrogen cycling regulates the cnidarian--algal symbiosis},
  author={R{\"a}decker, Nils and Escrig, St{\'e}phane and Spangenberg, Jorge E and Voolstra, Christian R and Meibom, Anders},
  journal={Nature Communications},
  volume={14},
  number={1},
  pages={6948},
  year={2023},
  publisher={Nature Publishing Group UK London}
}

@article{gorman2025stability,
  title={Stability of the cnidarian--dinoflagellate symbiosis is primarily determined by symbiont cell-cycle arrest},
  author={Gorman, Lucy M and Tivey, Trevor R and Raymond, Evan H and Ashley, Immy A and Oakley, Clinton A and Grossman, Arthur R and Weis, Virginia M and Davy, Simon K},
  journal={Proceedings of the National Academy of Sciences},
  volume={122},
  number={14},
  pages={e2412396122},
  year={2025},
  publisher={National Academy of Sciences}
}

@book{douglas2010symbiotic,
  title={The symbiotic habit},
  author={Douglas, Angela E},
  year={2010},
  publisher={Princeton University Press}
}

@article{douglas2015special,
  title={The special case of symbioses: mutualisms with persistent contact},
  author={Douglas, Angela E},
  journal={Mutualism},
  pages={20--34},
  year={2015},
  publisher={Oxford University Press Oxford, UK}
}

@article{bronstein2015study,
  title={The study of mutualism},
  author={Bronstein, Judith L},
  journal={Mutualism},
  pages={3--19},
  year={2015},
  publisher={Oxford University Press Oxford, UK}
}

@article{holland2015population,
  title={The population ecology of mutualism},
  author={Holland, Julian Nathaniel},
  journal={Mutualism},
  pages={133--158},
  year={2015},
  publisher={Oxford University Press Oxford, UK}
}

@article{falkowski1993population,
  title={Population control in symbiotic corals},
  author={Falkowski, Paul G and Dubinsky, Zvy and Muscatine, Leonard and McCloskey, Lawrence},
  journal={Bioscience},
  volume={43},
  number={9},
  pages={606--611},
  year={1993},
  publisher={JSTOR}
}

@article{radecker2021heat,
  title={Heat stress destabilizes symbiotic nutrient cycling in corals},
  author={R{\"a}decker, Nils and Pogoreutz, Claudia and Gegner, Hagen M and C{\'a}rdenas, Anny and Roth, Florian and Bougoure, Jeremy and Guagliardo, Paul and Wild, Christian and Pernice, Mathieu and Raina, Jean-Baptiste and others},
  journal={Proceedings of the national academy of sciences},
  volume={118},
  number={5},
  pages={e2022653118},
  year={2021},
  publisher={National Academy of Sciences}
}

@article{tanaka2018stoichiometry,
  title={The stoichiometry of coral-dinoflagellate symbiosis: carbon and nitrogen cycles are balanced in the recycling and double translocation system},
  author={Tanaka, Yasuaki and Suzuki, Atsushi and Sakai, Kazuhiko},
  journal={The ISME journal},
  volume={12},
  number={3},
  pages={860--868},
  year={2018},
  publisher={Oxford University Press}
}

@article{pernice2012single,
  title={A single-cell view of ammonium assimilation in coral--dinoflagellate symbiosis},
  author={Pernice, Mathieu and Meibom, Anders and Van Den Heuvel, Annamieke and Kopp, Christophe and Domart-Coulon, Isabelle and Hoegh-Guldberg, Ove and Dove, Sophie},
  journal={The ISME journal},
  volume={6},
  number={7},
  pages={1314--1324},
  year={2012},
  publisher={Oxford University Press}
}

@article{radecker2018using,
  title={Using Aiptasia as a model to study metabolic interactions in cnidarian-Symbiodinium symbioses},
  author={R{\"a}decker, Nils and Raina, Jean-Baptiste and Pernice, Mathieu and Perna, Gabriela and Guagliardo, Paul and Kilburn, Matt R and Aranda, Manuel and Voolstra, Christian R},
  journal={Frontiers in Physiology},
  volume={9},
  pages={214},
  year={2018},
  publisher={Frontiers Media SA}
}

@article{hale2021ecological,
  title={Ecological theory of mutualism: robust patterns of stability and thresholds in two-species population models},
  author={Hale, Kayla RS and Valdovinos, Fernanda S},
  journal={Ecology and Evolution},
  volume={11},
  number={24},
  pages={17651--17671},
  year={2021},
  publisher={Wiley Online Library}
}

@article{sachs2006pathways,
  title={Pathways to mutualism breakdown},
  author={Sachs, Joel L and Simms, Ellen L},
  journal={Trends in ecology \& evolution},
  volume={21},
  number={10},
  pages={585--592},
  year={2006},
  publisher={Elsevier}
}

@article{becks2025emergent,
  title={Emergent feedback between symbiosis form and population dynamics},
  author={Becks, Lutz and Gaedke, Ursula and Klauschies, Toni},
  journal={Trends in Ecology \& Evolution},
  volume={40},
  number={5},
  pages={449--459},
  year={2025},
  publisher={Elsevier}
}

@article{cunning2014not,
  title={Not just who, but how many: the importance of partner abundance in reef coral symbioses},
  author={Cunning, Ross and Baker, Andrew C},
  journal={Frontiers in microbiology},
  volume={5},
  pages={400},
  year={2014},
  publisher={Frontiers Media SA}
}

@article{xiang2020symbiont,
  title={Symbiont population control by host-symbiont metabolic interaction in Symbiodiniaceae-cnidarian associations},
  author={Xiang, Tingting and Lehnert, Erik and Jinkerson, Robert E and Clowez, Sophie and Kim, Rick G and DeNofrio, Jan C and Pringle, John R and Grossman, Arthur R},
  journal={Nature communications},
  volume={11},
  number={1},
  pages={108},
  year={2020},
  publisher={Nature Publishing Group UK London}
}

@article{yellowlees2008metabolic,
  title={Metabolic interactions between algal symbionts and invertebrate hosts},
  author={Yellowlees, David and Rees, T Alwyn V and Leggat, William},
  journal={Plant, cell \& environment},
  volume={31},
  number={5},
  pages={679--694},
  year={2008},
  publisher={Wiley Online Library}
}

@article{cui2023carbon,
  title={A carbon-nitrogen negative feedback loop underlies the repeated evolution of cnidarian--Symbiodiniaceae symbioses},
  author={Cui, Guoxin and Mi, Jianing and Moret, Alessandro and Menzies, Jessica and Zhong, Huawen and Li, Angus and Hung, Shiou-Han and Al-Babili, Salim and Aranda, Manuel},
  journal={Nature Communications},
  volume={14},
  number={1},
  pages={6949},
  year={2023},
  publisher={Nature Publishing Group UK London}
}

@article{cunning2017dynamic,
  title={A dynamic bioenergetic model for coral-Symbiodinium symbioses and coral bleaching as an alternate stable state},
  author={Cunning, Ross and Muller, Erik B and Gates, Ruth D and Nisbet, Roger M},
  journal={Journal of Theoretical Biology},
  volume={431},
  pages={49--62},
  year={2017},
  publisher={Elsevier}
}

@article{muller2009dynamic,
  title={Dynamic energy budgets in syntrophic symbiotic relationships between heterotrophic hosts and photoautotrophic symbionts},
  author={Muller, Erik B and Kooijman, Sebastiaan ALM and Edmunds, Peter J and Doyle, Francis J and Nisbet, Roger M},
  journal={Journal of Theoretical Biology},
  volume={259},
  number={1},
  pages={44--57},
  year={2009},
  publisher={Elsevier}
}

@article{gross2006generalized,
  title={Generalized models as a universal approach to the analysis of nonlinear dynamical systems},
  author={Gross, Thilo and Feudel, Ulrike},
  journal={Physical Review E—Statistical, Nonlinear, and Soft Matter Physics},
  volume={73},
  number={1},
  pages={016205},
  year={2006},
  publisher={APS}
}

@article{johnson2013competition,
  title={Competition for benefits can promote the persistence of mutualistic interactions},
  author={Johnson, Christopher A and Amarasekare, Priyanga},
  journal={Journal of Theoretical Biology},
  volume={328},
  pages={54--64},
  year={2013},
  publisher={Elsevier}
}

@article{jones2012fundamental,
  title={The fundamental role of competition in the ecology and evolution of mutualisms},
  author={Jones, Emily I and Bronstein, Judith L and Ferri{\`e}re, R{\'e}gis},
  journal={Annals of the New York Academy of Sciences},
  volume={1256},
  number={1},
  pages={66--88},
  year={2012},
  publisher={Wiley Online Library}
}

@article{may1976models,
  title={Models for two interacting populations},
  author={May, Robert M},
  journal={Theoretical ecology: principles and applications},
  pages={49--70},
  year={1976},
  publisher={Blackwell Scientific}
}

@article{cui2023molecular,
author = {Guoxin Cui  and Migle K. Konciute  and Lorraine Ling  and Luke Esau  and Jean-Baptiste Raina  and Baoda Han  and Octavio R. Salazar  and Jason S. Presnell  and Nils Rädecker  and Huawen Zhong  and Jessica Menzies  and Phillip A. Cleves  and Yi Jin Liew  and Cory J. Krediet  and Val Sawiccy  and Maha J. Cziesielski  and Paul Guagliardo  and Jeremy Bougoure  and Mathieu Pernice  and Heribert Hirt  and Christian R. Voolstra  and Virginia M. Weis  and John R. Pringle  and Manuel Aranda },
title = {Molecular insights into the Darwin paradox of coral reefs from the sea anemone Aiptasia},
journal = {Science Advances},
volume = {9},
number = {11},
pages = {eadf7108},
year = {2023},
doi = {10.1126/sciadv.adf7108},
URL = {https://www.science.org/doi/abs/10.1126/sciadv.adf7108},
eprint = {https://www.science.org/doi/pdf/10.1126/sciadv.adf7108}
}

@article{dimond2008symbiosis,
  title={Symbiosis regulation in a facultatively symbiotic temperate coral: zooxanthellae division and expulsion},
  author={Dimond, J and Carrington, E},
  journal={Coral Reefs},
  volume={27},
  number={3},
  pages={601--604},
  year={2008},
  publisher={Springer}
}

@article{kayal2018phylogenomics,
  title={Phylogenomics provides a robust topology of the major cnidarian lineages and insights on the origins of key organismal traits},
  author={Kayal, Ehsan and Bentlage, Bastian and Sabrina Pankey, M and Ohdera, Aki H and Medina, Monica and Plachetzki, David C and Collins, Allen G and Ryan, Joseph F},
  journal={BMC evolutionary biology},
  volume={18},
  number={1},
  pages={68},
  year={2018},
  publisher={Springer}
}

@article{lajeunesse2018systematic,
  title={Systematic revision of Symbiodiniaceae highlights the antiquity and diversity of coral endosymbionts},
  author={LaJeunesse, Todd C and Parkinson, John Everett and Gabrielson, Paul W and Jeong, Hae Jin and Reimer, James Davis and Voolstra, Christian R and Santos, Scott R},
  journal={Current biology},
  volume={28},
  number={16},
  pages={2570--2580},
  year={2018},
  publisher={Elsevier}
}

@article{massing2025generalized,
  title={Generalized dynamics of cooperating bacteria},
  author={Massing, Jana C and Gross, Thilo and Yeakel, Justin D and Fahimipour, Ashkaan K},
  journal={bioRxiv},
  pages={2025--05},
  year={2025},
  publisher={Cold Spring Harbor Laboratory}
}

@book{guckenheimer2013nonlinear,
  title={Nonlinear oscillations, dynamical systems, and bifurcations of vector fields},
  author={Guckenheimer, John and Holmes, Philip},
  year={2013},
  publisher={Springer Science \& Business Media}
}

@article{guckenheimer1997computing,
  title={Computing hopf bifurcations i},
  author={Guckenheimer, John and Myers, Mark and Sturmfels, Bernd},
  journal={SIAM Journal on Numerical Analysis},
  volume={34},
  number={1},
  pages={1--21},
  year={1997},
  publisher={SIAM}
}

@article{stiefs2008computation,
  title={Computation and visualization of bifurcation surfaces},
  author={Stiefs, Dirk and Gross, Thilo and Steuer, Ralf and Feudel, Ulrike},
  journal={International Journal of Bifurcation and chaos},
  volume={18},
  number={08},
  pages={2191--2206},
  year={2008},
  publisher={World Scientific}
}

\end{document}